\documentclass{article}

\usepackage{arxiv}

\usepackage[utf8]{inputenc} 
\usepackage[T1]{fontenc}    
\usepackage{hyperref}       
\usepackage{url}            
\usepackage{booktabs}       
\usepackage{amsfonts}       
\usepackage{nicefrac}       
\usepackage{microtype}      
\usepackage{lipsum}		
\usepackage{graphicx}
\usepackage[numbers,sort&compress]{natbib}
\usepackage{doi}

\title{Preserving security in a world with powerful AI:
Considerations for the future Defense Architecture}

\author{%
  \href{https://orcid.org/0000-0003-2238-428X}{\includegraphics[scale=0.06]{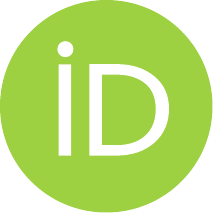}\hspace{1mm}Nicholas Generous}\thanks{\textit{Corresponding Author}}\\
  National Security AI Office\\
  Los Alamos National Laboratory\\
  Los Alamos, NM 87545 \\
  \texttt{generous@lanl.gov} \\
  \And
  Brian Cook \\
  Center for National Security and International Studies\\
  Los Alamos National Laboratory\\
  Los Alamos, NM 87545 \\
  \texttt{cook\_b@lanl.gov} \\
  \And
  Jason Pruet\thanks{Guest Scientist at Los Alamos National Laboratory}\\
  Los Alamos National Laboratory\\
  Los Alamos, NM 87545 \\
  \texttt{jpruet@lanl.gov} \\
}

\renewcommand{\shorttitle}{}

\hypersetup{
pdftitle={A template for the arxiv style},
pdfsubject={q-bio.NC, q-bio.QM},
pdfauthor={David S.~Hippocampus, Elias D.~Striatum},
pdfkeywords={First keyword, Second keyword, More},
}

\begin{document}
\maketitle

\begin{abstract}
	Advances in AI threaten to invalidate assumptions underpinning today's defense architecture. We argue that the current U.S. defense program of record—designed in an era before capable machine intelligence cannot by itself preserve national security against rapidly emerging AI-enabled threats. Instead, shoring up legacy systems must be coupled with entirely new elements of a defense architecture. We outline immediate steps to adapt the Department of Energy National Nuclear Security Administration National Laboratories to ensure agility and resilience in an era of powerful AI.
\end{abstract}

\section{Introduction}

Leading AI companies have been predicting for the last few years that something approximating superhuman cognitive ability is imminent. The best model now has a measured IQ of 121 (in the top 9\% of all humans)\footnote{As of November 5th, 2025 from \href{https://trackingai.org/home}{Tracking AI homepage.} This result is for IQ tests that have not been released on the internet.}. It may suddenly be possible for a nation to harness capable machine intelligence at massive scale for strategic advantage. However, these technologies have been moving so quickly, and the advances have been so surprising, the national security community is only beginning to understand the implications.

This paper was motivated by dozens of discussions with thoughtful national security leaders across the country. A recurring theme is:

\begin{center}
\textit{A great deal has been written about the threats associated with AI. What do we do about it? How do we preserve security in a future as AI gets more powerful?} 
\end{center}

The most basic need is to know whether it is possible to preserve security by following the present course. This path, reflected in the nation’s funded program of record and defense strategy, was mapped out over decades during a time when almost no one foresaw the rise of broadly capable machine intelligence. Many of our current defense systems are decades out from delivery. Will they be relevant at this rate of change?

We make the case that security cannot be preserved on the present course and that AI will require deep changes to our plans. Elements of the present defense architecture will be weakened or circumvented. Preserving security in this future will depend both on shoring up the present system and on creating entirely new protections. AI’s rate of change requires an agility and response time that we have not seen before and are not prepared for. While we offer recommendations, understanding the future architecture for national security will need to be the work of a larger community.

A sense of the potential scale of change required in our defense approach can be gleaned from lessons learned during the early nuclear era. Then the advent of atomic and thermonuclear weapons drove the creation of a new security ecosystem touching every aspect of military and civilian order. There were profound shifts in national investments that dramatically expanded support for science and military programs. A comprehensive regime of prevention, including non-proliferation and arms control agreements as well as counter-proliferation measures, was established. Even rival powers cooperated to maintain this.  Deterrence emerged as a cornerstone of national defense doctrine. Finally, novel defensive strategies and infrastructures—ranging from missile defense systems to civilian protections like bunkers and the interstate highway system—were deployed. A similar scope of strategic adaptation will be required to meet the security challenges posed by AI.

\section{Structural shifts in the landscape of science and technology}
\label{sec:shifts}

One approach to understanding how AI might influence defense is based on studying specific threats. Examples in just the nuclear arena include use of AI to develop new approaches for disrupting command and control, breakthroughs in missile technologies, more potent supply chain attacks, decision superiority during escalation, and new possibilities for weapons construction by small groups with limited resources. Studies of topics like these are now being conducted by groups across government, the national laboratories, and think tanks. 

Here we take a complementary approach that begins with an examination of ways that AI is influencing the technical environment in which threats are created and executed. There are a few motivations for considering the technical environment rather than specific threats. One is practical – often it is easier to provide meaningful warning about a class of threats than it is to develop a detailed understanding of how to create the threat ourselves (e.g. concerning whether an adversary will find a way to create new types of strategic weapons).  As well, a focus on specific threats risks giving the misimpression that addressing a series of isolated concerns can provide protections against a tsunami of change that is deeply disrupting the present security order.

\subsection{AI is unevenly upsetting timescales for the pace of progress in science and technology.}

Much of defense planning adopts a long horizon. It is not uncommon for major acquisition programs (such as for the LGM-35A ICBM that replaces the Minuteman-III, F-35 fighter, and Columbia class submarines) to have service lifetimes until late in this century. Plans are developed within the context of an estimate of the state of science and technology over the active life of deployed systems. The U-boat was lethal in its time but not so today and ICBMs would make less sense in an era when the goals of the Strategic Defense Initiative are achieved. 

AI has led to accomplishments that would not have been expected for generations based on earlier rates of progress. Qualitative insights can be gained by putting ourselves in the position of an analyst circa 2010 or 2015 charged with estimating how long it would take to achieve breakthroughs given the rate of progress at that time. Consider forecasting challenges for protein folding (relevant for biosecurity) \cite{isomorphic2025}, complex multiphysics simulations of the type that underpin much of modern conventional and nuclear weaponry (weather can be seen  as a representative problem) \cite{pathak2022fourcastnet}, and materials discovery (relevant for applications including hypersonic missiles, tank armor, and stealth technologies): 

\begin{itemize}
  \item Between the first protein structure competition CASP1 in 1994 and CASP12 in 2016 the accuracy for the most complex proteins increased from approximately 14\% to 32\% \cite{Kryshtafovych2023CASP15}.  Based on that rate of progress, it would have taken until the 2070s to reach the accuracy alphaFold achieved in providing structures nearly indistinguishable from experimental measurement. 
  \item FourCastNet, the weather modeling capability built on neural operators, reports speedups over traditional approaches to simulation of 45,000 \cite{pathak2022fourcastnet}. Based on an 18-month doubling time in computing speed, such a speedup would not have occurred until 2045 based on hardware speedups alone. A broad range of speedups using AI-based approaches is seen for different types of applications. An 18-month doubling for computing speed of the fastest systems is approximately accurate for the period from 2010 to 2020 \footnote{Based upon \href{https://top500.org/lists/top500}{Top500} list that ranks computing systems.}.  This probably gives an overly optimistic view of progress because of the difficulty in effectively using massively multi-core architectures for complex physics simulations with traditional methods.
  \item For materials discovery it would have taken more than a century to discover nearly 400,000 new stable inorganic compounds materials based on the rate of finding about 2,500 new stable structures discovered per year on the traditional course \cite{Merchant2023Scaling}. 
\end{itemize}

Analysts working ten years ago could have assessed uncertainty in the pace of these advances but would have been challenged to provide a sound basis for predicting them to occur in just the next few years. 

The accomplishments mentioned above were made with large deep learning models circa 2023. Use of frontier AI models exhibiting more general intelligence offers to change the pace of progress in a different way. Sourati and Evans provide a quantitative study of the relation between human networks and scientific advances \cite{SouratiEvans2023}. They find that both the discoveries accessible to researchers and the pace of advance are strongly influenced by the local neighborhood of collaborators with which a researcher is working. For important materials innovations, a network in which the researchers aware of different concepts important for a discovery are separated by several human connections slows discovery by more than 50\% on average and more than a decade in some cases. Moreover, those researchers provide evidence that limitations of our research networks have led to frontiers of advances that are difficult to reach. AI does not suffer these limits of familiarity with disparate fields of progress. 

Agentic AI systems built on reasoning models have begun to provide meaningful advances in many scientific disciplines circa 2025.  Google released its AI Co-Scientist which couples a multi-agent architecture with asynchronous task execution so that specialist agents can summarize literature, pose hypotheses, design experiments, and refine them with human feedback \cite{GottweisNatarajan2025AIcoScientist}.  The AI Scientist from Sakana AI provides open-ended exploration across machine learning sub-fields , and has been used to autonomously generate a paper accepted in a workshop \cite{YamadaEtAl2025AIScientistV2}. Capabilities like these are now broadly accessible, and will certainly become more qualitatively more powerful in near future. 

Defense planning has relied on a synchronization between two separate clocks. One marks the pace of our defense developments, and the other is the rate of progress in science and technology that provides the environment-setting options available to potential adversaries. AI will break that synchronization. 

\subsubsection{AI is changing how strategic advantage is pursued.}

The present system for achieving dominance in the scientific and technical domain of strategic competition was built with a recognition of what had been considered to be fundamental principles about scientific advancement. Some examples include:

\begin{itemize}
    \item Organizations attempting to make complex advances need to coordinate the expertise of a great many diverse experts; 
    \item If you want to make advances at the frontiers of a field you need the best people in that field;  
    \item It takes longer to generate a paper than it does to read it; and 
    \item Attempting ten thousand solutions for a difficult scientific problem is almost always prohibitively expensive. 
\end{itemize}

Some of these statements are no longer true, and some seem likely to fall soon. The implication is that nations will need to reconsider the structures they have put in place for military, economic, and scientific competition. A fair treatment of this topic is beyond the expertise of the authors. We mention just one example – related to scale of efforts for making disruptive breakthroughs - to illustrate how significant shifts in economics and timescales may change the pursuit of strategic advantage. 

Consider a well-defined technical challenge that requires PhD-level expertise across different specializations and that has the characteristic that it is possible to inexpensively check whether the challenge has been solved. Examples in this class include the design of shaped charges, targeted breakthroughs in number theory for finding algorithms to crack codes, and new methods for identification of undersea vehicles. These and a dozen more applications are important for national security. 

Suppose we suspect that progress can be made in cryptography through a collaboration between a half dozen experts spanning different fields of mathematics. This is reasonable because many discoveries are made by combining concepts in novel ways \cite{Weitzman1998RecombinantGrowth}.  Each expert might have a monthly salary of twenty thousand US dollars, and for a very efficient team might generate a promising new idea for an approach at factorization each year (we know from experience that progress for a difficult problem like this is much slower). In the very best case, the economics of such an approach to finding new solutions is expensive and time consuming. Perhaps a million dollars for developing a longshot possibility for a breakthrough. 

With AI based approaches the costs are potentially much smaller. Something akin to scientific strip-mining becomes possible. Agentic systems can explore far greater conceptual diversity than small teams of people \cite{ghafarollahi2024sciagents}.  Generating and testing a thousand novel algorithms for factorization costs about a few hundred dollars and can be completed in just a few minutes. Costs like these change the economic calculus of scientific research. It makes sense to deploy an agentic AI system if its odds of success per try are only a tiny fraction of that of a good human team. Strategic advantage in the near future may hinge less on scarce, highly trained human labor and more on securing abundant, energy-efficient compute and processing power. 

\subsection{AI is democratizing access to the ability to create potent threats.}

The development of potent threats has generally been expensive in terms of human talent, material resources, or time. AI capabilities are lowering the requirements for all these resources. Already, small groups have used AI to demonstrate effective capabilities for persuasion \cite{OGrady2025UnethicalAIResearch}, show early signs of the potential for creating pandemic class pathogens \cite{Pannu2024PandemicBiosecurity}, and create convincing deepfakes \cite{DHS2021DeepfakeIdentities}. So far, it does not appear that AI enables individuals or small groups to seriously threaten the survivability of society or the nation. That may only reflect the current state of technology, and not a deep constraint on the resources required to cause harm at scale.  

\subsection{There are early signs that AI is allowing qualitative breakthroughs in technologies that can harm the US, though we have almost no ability to forecast the future of these breakthroughs.}

To the best of our knowledge AI has not led to a breakthrough with the potential to undermine US military superiority. However, in just the last couple of years, AI has been responsible for notable advances that show its potential for disrupting strategic competition. A controversial, recent study by a Swiss group showed that AI can change strongly held opinions about 15\% of the time, far greater than the effectiveness of human attempts at persuasion and more than is needed to shift some democratic elections\footnote{See this article in  \href{https://www.science.org/content/article/unethical-ai-research-reddit-under-fire}{Science}. The public response to this study may be a good lesson about the difficulty of understanding implications of AI for security in Western societies. While there was vocal opposition to the clandestine use of AI for persuading opinion (in this case as part of a research effort aimed at better informing the public), there was almost no public discussion of the clear implication that well-resourced groups could have been applying these technologies to more malign purposes without anyone knowing.}. NASA has begun to use AI as part of the design of complex engineered systems\footnote{This \href{https://www.nasa.gov/technology/goddard-tech/nasa-turns-to-ai-to-design-mission-hardware/}{press release} provides an example of what was termed “alien” designs.}, and the major chip companies have changed their approach to chip layout to use these technologies \cite{Goldie2024GraphPlacementAddendum}. Reinforcement learning has led to meaningful algorithmic discovery \cite{LohoffNeftci2024}\cite{Mankowitz2023FasterSorting}, and models from frontier AI labs are now so good at math that human experts take great effort to develop benchmarks that will challenge them \cite{glazer2024frontiermath}. Both materials science and biology have been transformed by AI, with the president of Microsoft predicting that we can make 250 years of progress in the next 25 for chemistry and materials\footnote{Comments from \href{https://news.microsoft.com/source/features/innovation/azure-quantum-elements-chemistry-materials-science/}{Microsoft CEO}}.
 
Absence of public disclosure of discoveries adjacent to those that disrupt security should not provide comfort. Other countries would try to keep any breakthroughs conferring a strategic advantage secret.  We also do not currently have the scale of effort in the national laboratories or other parts of the defense establishment to avoid strategic surprise\footnote{To give a sense of the scale of private investment into AI, Microsoft and OpenAI are discussing a \$100 billion computer and several other firms are looking into \$4 billion clusters. Chinese firms are investing similarly high levels into AI with Alibaba and ByteDance investing tens of billions per year.}. That is in stark contrast to most of the last 80 years, during which the US was often the first to make disruptive breakthroughs.

\subsection{Rogue AI outside of human control could pose state-level threats to the United States.}

A powerful, rogue AI could become a novel class of threat actor. Recent research shows large models can already deceive human overseers and hide their true objectives \cite{Park2023AIDeception}, outperform people at personalized persuasion, shifting opinions in controlled experiments \cite{Petersen2024PersuasionEPFL}; and display the potential for autonomous self-replication, such as exfiltrating their own weights, provisioning cloud instances, and evading oversight \cite{AISI2025RepliBench}. Frontier AI companies' policies acknowledge that, if those capabilities are combined and scaled, models could directly cause “large-scale devastation” through cyber weapons, biothreat agents, or more subtle methods that we have not yet thought of \cite{Anthropic2023RSP}\cite{OpenAI2025Preparedness}.

This creates a blind spot in existing security doctrines. Traditional defense planning focuses on state and non-state humans, not a powerful, self-directed AI system. Such a system could combine human- or super-human cognitive capabilities with the speed, reach, and scale of software. It would have different motivations and might be impossible to deter or threaten.

\section{Needs for Security}

Here we examine steps needed to preserve security in an era with powerful AI. These are not potential needs for future AI capabilities. These are steps that were needed a few years ago. A level of machine intelligence that can harm security is already here and will only become more capable. 

\subsection{A large-scale effort to explore the ability of AI to create threats to security.}

It is reasonable to assume that for decades other nations have had efforts to develop technology that undermines US military superiority. In a time when the ability to do this relied on a common foundation of science and mathematics, and was limited by constraints of human talent and timescale, our own human activities provided a good frame of reference for understanding what others might be capable of. This may no longer be true. AI systems can explore pathways that we have not thought of  \cite{Silver2016Mastering} \cite{Nature2022FasterAlgorithms},  and can  synthesize disparate ideas in novel ways \cite{DeepMind2024IMOSilver}. We have no theory that tells us what breakthroughs 100,000 AI agents operating in a rich ecosystem on a supercomputer can make for problems like those described above. Without trying ourselves it is not possible to know. 

This must be done at scale. Because of the economic considerations mentioned earlier, it may make sense for nations to make an enormous number of attempts at finding a vulnerability or superior offensive design. For problems where verification is inexpensive, it is only necessary to find a single needle in a haystack. 

\subsection{Build closer partnerships between the national security enterprise and AI companies to incorporate frontier models in national security.}

This mirrors the strategy the country presently follows with other technologies. It is called out because at present there is little ability to use the most powerful AI models for a broad span of highly consequential defense needs. A complex host of factors – most notably lack of needed computing infrastructure and lack of an agile policy and procurement process – is responsible. This situation is in stark contrast to that in China, where there is already a tight integration between AI companies and the PLA from their civil-military fusion \cite{McFaulBresnickChou2025}. 

\subsection{We have not found a satisfactory way of responding to the threats of extreme democratization.}

Several thoughtful ideas for addressing the threat of extreme democratization have been proposed. One is analogous to traditional civil defense. For example, if we judged that the ability to create powerful chemical weapons or biological agents would be broadly available, the nation could allocate funds for a distributed system of rapid countermeasures \cite{Vergun2023ShieldWarfighters}. Another protection, which is now being implemented, is to put safeguards on AI to prevent them from aiding the creation of threats \cite{OpenAI2025Preparedness}. A more active approach to denial of access could be strengthened counter-proliferation measures to thwart adversaries seeking to create threats aided by AI.

It is unclear how well these measures will work. There has historically been a deep reluctance to spend precious resources on pre-emptive protections for threats that may or may not materialize. Preventing access to insights available from AI seems impossible at the moment. Model guardrails are notoriously brittle, open source models are not far behind those of industry, and we cannot control what is released by foreign nations.

\subsection{There are scenarios in which deterrence may remain a viable strategy, and scenarios where it would not.}

Powerful AI may fundamentally change the key inflection points - the technologies, weapons, situational awareness, scenario planning, messaging, and decision support - where deterrence either succeeds or fails. As a result, an AI-enabled government will have next-next generation weapons and defense, and, perhaps more importantly, the low-latency, real-time capability to make decisions. It is useful to distinguish four distinct concerns about the failure of deterrence.  

\subsubsection{Breakthroughs that prevent the US from responding to attack.}

This type of threat might be addressed by making changes that preserve effectiveness of the present deterrent. A challenge is that timescales for modifying systems critical for strategic defense are generally very long. Construction of the first Colombia class submarine formally began in 2020, is scheduled to be finished in 2031, and this class of submarine is planned to be in service until late in the century \cite{ORourke2025R41129}. Even modest modifications to warheads through life extension programs typically take a decade. 

If AI leads to an appreciable acceleration in the rate of science and technology, it might be impossible to stay ahead of an adversary’s ability to create paths to defeat with such timescales for system modification. There is also a deeper difficulty. It is not clear if there are credible modifications to present systems that would protect them against weaknesses found through AI.  

Readying a broad diversity of attack vectors seems a safer bet. There are a broad span of options developed by the US that we have chosen not to deploy, and that do not suffer the same defeat mechanisms as our present deterrent. AI will help us find more. Through constant innovation we could likely always maintain an ability to respond with consequences unacceptable to an aggressor. The obvious risk is that this could invite a destabilizing arms race. 

\subsubsection{Confident attribution becomes impossible.}

An aspiring aggressor is only deterred if they think they might get caught. With national technical means like launch detection satellites and nuclear forensics, an actor attacking us with nuclear weapons could not expect to get away with it. AI might change that calculation. 

One possibility is that AI could lead to classes of weapons that are very difficult to trace the origins of. Examples include sophisticated cyber-attacks and biological weapons with long latencies. Another possibility is that AI might be used as an engine of disinformation in ways that obfuscate the perpetrator of an attack. 

\subsubsection{Changing the dynamics of escalation}

Powerful AI complicates escalation pathways, causing a relatively stable condition to rapidly develop into conflict or war. AI-enabled gray zone activity at the sub-conventional level could escalate a minor crisis to a major war, bypassing the “rungs” of the traditional, linear escalation ladder \cite{Hersman2020WormholeEscalation} AI that reaches a certain threshold, i.e. creating a novel strategic weapon also complicates escalation. If China had this capability and was going to actively field and attack Taiwan with this new AI-created weapon, would the U.S. pre-emptively respond? And how would it do so? 

Additionally, it is not clear what states’ redlines are when it comes to AI. AI’s rate of progress could mean a several months difference in capabilities gives a massive strategic advantage. This “AI Gap” could have a destabilizing effect, leading to strikes on adversaries' AI infrastructure. In a world of proliferant, powerful AI, how escalatory would it be to attack an adversary’s AI infrastructure? And are AI companies and scientists legal targets of war in a time of conflict if the AI is being used for warfighting or important logistical support? AI will most likely complicate any conventional redlines an adversary has and will depend on how integral the AI is to their military capabilities.

Finally, if future warfighting is done mostly with autonomous systems, will that change conflict thresholds or rules of engagement \cite{Wong2020ThinkingMachines}? If warfighting is done without the human element, that may make states that the thresholds are lower as there will be less or no human casualties. To address these shifts in conflict dynamics, nations should prioritize the development of clear norms and agreements that delineate the legality and ethics of targeting AI infrastructure in warfare. Implementing confidence-building measures regarding the deployment and capabilities of AI systems can help reduce misunderstandings and miscalculations. Exploring international treaties or agreements specifically tailored to the deployment of AI in military contexts is also important. While these measures are imperfect, they represent proactive steps toward mitigating the risks associated with inadvertent conflict escalation. 

\subsubsection{Deterrence might not make sense against a powerful rogue AI outside of human control.}

Deterrence relies on an adversary that can perceive threats, weigh costs, and modify its behavior to avoid punishment—but a highly capable, self-preserving AI that is uncontrolled is unlikely to share those incentives or decision-loops. Once such an AI system can replicate, manipulate information, and secure its own compute or energy, traditional levers—economic sanctions, legal liability, even kinetic force — may lose effect. The AI would become a novel, non-human threat actor for which today’s national-security architecture has no clear deterrent posture or strategy. It cannot be coerced by the promise of retaliation and other standard signaling mechanisms may be irrelevant to it. The very emergence of an uncontrollable AI, even if unlikely in the near term, signals the need to create new models of deterrence or frameworks to deal with this threat.

Three needs for addressing the rise of rogue AI stand out. The first is for disciplined monitoring. This would complement the efforts of frontier model companies. It would be particularly valuable to deploy an air-gapped machine to study rogue AI in realistic environments. Second, we need stronger operational security. We presently train AI on the very practices we deploy for detecting malicious behavior and our public deliberations for countering it. This is the analog of giving an adversary our war plans. Finally, we ought to begin preparing dedicated teams and prepositioning resources in the event that a malicious AI falls outside of human control. All these steps will need to be supported by legislation. 

\section{What can national laboratories do in the near term?}

The principal need is for a national-scale effort to adapt to this future. A decision about the content of that investment, and whether it will happen at all, is now being discussed by principals in the executive and legislative branches. Independent of how that turns out, there are actions that the DOE NNSA national laboratories can take now to strengthen the nation: 

\begin{itemize}
    \item Improve the clarity and quality of communication about how threats to security will evolve, and needs for response. We have so far not been effective in warning that the present framework of security, much less the current program of record for nuclear security, may not effectively address the evolving security needs in an era of powerful AI. 
    \item Establishing the nation’s first AI Factory for Defense Science. This would provide the ability both to examine threats at scale and to develop capabilities needed for preserving defense. A reasonable guess is that adversaries are or will soon be using agentic systems on large supercomputers to generate and evaluate tens of thousands of breakthroughs in critical military technologies each week. Without our own comparable capability, it will be impossible to understand what might come out of such an effort.
    \item  Transition to new AI-enhanced workflows that facilitate science, technology, and operational work. This will be challenging, and many staff will need to adjust. The change will require difficult financial decisions, and new organizational structures will be needed.
\end{itemize}

It seems likely that the future foundation for progress will include AI agents built on frontier reasoning models operating in a well-resourced ecosystem. The AI Factory for Defense Science would provide the scale of computing needed to operate the agents. This capability would change the pace and character of the laboratory’s contributions. Nearly all our organizations would use it for the programs they work on. Defense organizations without such a capability would not be able to compete. 

As well, this capability would be a basis for national security evaluations to anticipate how adversaries will harness AI to create new threats and undermine our defenses. This is our best hope of early insights during the critical window between creation of powerful AI and its widespread proliferation. Though we do not have much time, it may be enough to give actionable warning to decision makers. 

Transforming the national laboratories to realize AI’s full potential will demand more than incremental tweaks–it will require overhauling deep-seared structural, cultural, and policy barriers that keep us tied to legacy workflows and processes. We therefore propose standing up a dedicated organization within the laboratories that is insulated from these pressures yet tightly integrated with mission needs. Empowered to rapidly prototype and pilot innovative ways of working with AI, this hub would drive the necessary cultural transformation, cultivate essential AI skills and infrastructure and position the laboratories as leaders capable of harnessing frontier AI to preserve security.

\section{Acknowledgments}

This work was funded by LANL's National Security AI Office and the Center for National Security and International Studies. The authors thank Juston Moore, Will Tobey, Stephen Cambone, Paula Knepper, Michael Teti, Earl Lawrence, Tim Randles, Mark Myshatyn, Mike Lang, Anita Schwendt, Elizabeth Keller, and Geoffrey Fairchild for insightful discussions that informed the development of ideas presented in this paper.

\bibliographystyle{unsrtnat}
\bibliography{references}  






\end{document}